\documentclass[prb,twocolumn,amsmath,mathtools,a4paper,showpacs]{revtex4-1}
\usepackage{graphicx}
\usepackage{amsmath}

\usepackage{epsfig}
\usepackage{charter}
\usepackage{amssymb}
\usepackage{setspace}
\usepackage{amsmath}
\usepackage{csquotes}
\usepackage{float}
\usepackage {braket}
\usepackage{epstopdf}
\usepackage{subfigure}
\usepackage{color}

\usepackage[version=3]{mhchem}
\usepackage{mathtools}
\usepackage{gensymb}
\usepackage{hhline}
\usepackage{gensymb}
\usepackage{graphics}

\begin{document}
	
\title{Coexistence of Magnetism and Superconductivity in Separate Layers of Iron-Based Superconductor Li$_{1-x}$Fe$_x$(OH)Fe$_{1-y}$Se}
\author{C.\ V.\ Topping}
\affiliation{University of Oxford, Department of Physics, Clarendon Laboratory, Parks Road, Oxford OX1 3PU, United Kingdom}
\author{F.\ K.\ K.\ Kirschner}
\affiliation{University of Oxford, Department of Physics, Clarendon Laboratory, Parks Road, Oxford OX1 3PU, United Kingdom}
\author{S.\ J.\ Blundell}
\affiliation{University of Oxford, Department of Physics, Clarendon Laboratory, Parks Road, Oxford OX1 3PU, United Kingdom}
\author{P.\ J.\ Baker}
\affiliation{ISIS Facility, STFC Rutherford Appleton Laboratory, Didcot OX11 0QX, United Kingdom}
\author{D.\ N.\ Woodruff}
\affiliation{University of Oxford, Department of Chemistry, Inorganic Chemistry Laboratory, South Parks Road, Oxford OX1 3PU, United Kingdom}
\author{F.\ Schild}
\affiliation{University of Oxford, Department of Chemistry, Inorganic Chemistry Laboratory, South Parks Road, Oxford OX1 3PU, United Kingdom}
\author{H.\ Sun}
\affiliation{University of Oxford, Department of Chemistry, Inorganic Chemistry Laboratory, South Parks Road, Oxford OX1 3PU, United Kingdom}
\author{S.\ J.\ Clarke}
\affiliation{University of Oxford, Department of Chemistry, Inorganic Chemistry Laboratory, South Parks Road, Oxford OX1 3PU, United Kingdom}
\date{\today}

\begin{abstract}
The magnetic properties attributed to the hydroxide layer of
Li$_{1-x}$Fe$_x$(OH)Fe$_{1-y}$Se have been elucidated by the study of
superconducting and non-superconducting members of this family of
compounds.  Both a.c. magnetometry and muon spin relaxation
measurements of non-superconductors find a magnetic state existing
below $\approx 10~\rm K$ which exhibits slow relaxation of
magnetisation.  This magnetic state is accompanied by a low
temperature heat capacity anomaly present in both superconducting and
non-superconducting variants suggesting that the magnetism persists
into the superconducting state.  The estimated value of magnetic
moment present within the hydroxide layer supports a picture of a
glassy magnetic state, probably comprising clusters of iron ions of
varying cluster sizes distributed within the lithium hydroxide layer.
\end{abstract}

\maketitle

\section{Introduction}\label{Intro}

Iron selenides, as a subset of iron-based superconductors, have received substantial interest in recent years.  The parent compound, FeSe, was found to be a superconductor with $T_{\rm c}\approx 8~\rm K$\cite{Hsu2008,McQueen2009} which has a strong dependence on the precise stoichiometry.\cite{McQueen2009}  $T_{\rm c}$ may be enhanced with the application of hydrostatic pressure\cite{Margadonna2009,Medvedev2009} and by preparing FeSe as a single-layer.\cite{Wang2012,Liu2012,He2013} The enhancement of $T_{\rm c}$ due to pressure has been suggested to be due to changes in hybridization of Fe and Se orbitals\cite{Svitlyk2016} which may indicate electron doping from a Se$^{2-}$ to Fe$^{2+}$ charge transfer, while the enhanced $T_{\rm c}$ in monolayer FeSe is suggested to arise from electron doping either due to O-deficient substrate or Se loss to form FeSe$_{1-x}$.\cite{He2013}  Reduction of multilayer FeSe with K atoms also leads to the enhancement of superconducting $T_{\rm c}$.\cite{Miyata2015}  Chemical intercalation (the insertion of additional ions or molecules between FeSe layers) has also been employed to increase $T_{\rm c}$ with the intercalation of metal ions, A,  to form families of compounds, A$_x$Fe$_2$Se$_2$, reaching $T_{\rm c}$'s of approximately $30\-- 40~\rm K$ in bulk compounds A$_{0.8}$Fe$_{1.6}$Se$_{2}$.\cite{Guo2010,Wang2011,Krzton-Maziopa2011,Ying2012}  This process may be taken further by intercalating, along with the reducing alkali metal, larger molecules such as ammonia to form Li$_{x}$(NH$_2$)$_y$(NH$_3$)$_{1-y}$Fe$_2$Se$_2$ and pyridine to form Li$_x$(C$_5$H$_5$N)$_y$Fe$_{2-z}$Se$_2$ with superconducting transitions of $43~\rm K$\cite{Burrard-Lucas2012} and $45~\rm K$\cite{Krzton-Maziopa2012} respectively.  While this raising of $T_{\rm c}$ is intriguing and may be considered a result of increased interplanar spacing,\cite{Guterding2015} another important aspect is the new physical behaviour that is introduced by inserting additional material between the FeSe layers.

An important example is the system of layered hydroxide selenides, Li$_{1-x}$Fe$_x$(OH)Fe$_{1-y}$Se.\cite{Sun2015,Lu2014,Pachmayr2015}  These compounds consist of alternating layers of tetrahedral FeSe and Li$_{1-x}$Fe$_x$(OH) (shown in Fig. \ref{Struct}) and have the advantage over amine-intercalated compounds of being much more thermally stable.  The FeSe layers contain Fe vacancies whose concentration has a large effect on the superconducting properties.\cite{Sun2015,Woodruff2016}  The hydroxide layers always contain a roughly $4:1$ ratio of Li:Fe with a $+2$ oxidation state of iron.\cite{Pachmayr2015,Sun2015}  A density functional theory (DFT) study suggests that this Li$:$Fe ratio is favoured to avoid lattice mismatch between the hydroxide and FeSe layers and allow commensurate stacking.\cite{Chen2016}  In addition, the Fe in the hydroxide layer may play a role in enhancing $T_{\rm c}$ by increasing charge transfer of $x$ electrons into the FeSe layers providing a similar level of doping to that in the metal/ammonia intercalates.\cite{Chen2016,Liu2017}  It has been shown that a certain degree of control may be exercised over the superconductivity of these compounds.  A post synthetic lithiation may be performed which introduces further Li into the system to displace Fe from the hydroxide layer which in turn fills Fe vacancies in the FeSe layer.\cite{Sun2015}  Utilising this method, it has been shown that superconductivity in these compounds requires both near stoichiometric FeSe layers ($y < 0.05$) \textit{and} the reduction of Fe in this layer below the +2 oxidation state as facilitated by the hydroxide layer Fe.\cite{Sun2015, Woodruff2016}  Increased lithiation to form the stoichiometric Li(OH)FeSe was found to destroy superconductivity.\cite{Woodruff2016}  In addition, there have been reports of magnetic order in these compounds, the magnetism likely originating in the hydroxide layer, with both ferromagnetic\cite{Wu2015,Pachmayr2015} and antiferromagnetic\cite{Lu2014,Khasanov2016} order being proposed.  Hydroxide layer magnetism has also been suggested to couple to and influence magnetism in the FeSe layer and therefore play a role in superconductivity.\cite{Liu2017}  This has been likened to the interaction between interstitial Fe and FeTe$_{1-x}$Se$_x$ layers in Fe$_{1+y}$Te$_{1-x}$Se$_x$.\cite{Thampy2012}  We have recently suggested that the Fe moments in the hydroxide layer of Li$_{1-x}$Fe$_x$(OH)Fe$_{1-y}$Se show glassy, rather than ordered, character\cite{Woodruff2016} owing to a lack of evidence for long range order in powder neutron diffraction.\cite{Woodruff2016,Lynn2015}

In this paper, further investigation of the hydroxide layer magnetism is reported via both static and dynamic magnetometry, heat capacity and muon spin relaxation ($\mu$SR) measurements on powder samples of Li$_{1-x}$Fe$_x$(OH)Fe$_{1-y}$Se.  Previous work\cite{Sun2015,Woodruff2016} has elucidated the compositional requirements of superconductivity and this has allowed both superconducting and non-superconducting variants of these compounds to be examined in an effort to disentangle the hydroxide magnetism from superconductivity and confirm its existence regardless of the presence of the superconducting state.

\section{Experiment}

Five powder samples of Li$_{1-x}$Fe$_x$(OH)Fe$_{1-y}$Se (three non-superconductors and two superconductors) were synthesised  as detailed in Refs.~\onlinecite{Sun2015} and \onlinecite{Woodruff2016}.  They are henceforth referred to as NonSC1, NonSC2, NonSC3, SC1 and SC2.  Non-superconducting samples were synthesised hydrothermally while superconducting samples underwent a post-synthetic lithiation after hydrothermal synthesis to ``turn on superconductivity'' by generating near stoichiometric FeSe layers via reductive intercalation of Li.\cite{Sun2015}  The structure and composition of each compound were derived from X-ray powder diffraction on the I11 beamline at the Diamond Light Source (Rutherford Appleton Laboratory, UK) with additional neutron powder diffraction performed at the Institut Laue-Langevin (Grenoble, France) on NonSC3.  Magnetometry was performed using a Quantum Design magnetic property measurement system (MPMS-XL).  Heat capacity measurements were carried out utilising a Quantum Design physical property measurement system (PPMS) on pressed pellets of powders in $0$ and $11~\rm T$ static magnetic fields.  $\mu$SR experiments\cite{Blundell1999, Yaouanc2011} were performed using a \ce{He^3} cryostat inserted in the MuSR spectrometer at the ISIS pulsed muon facility (Rutherford Appleton Laboratory, UK).\cite{King2013} Zero-field (ZF) measurements were performed in which the Earth's magnetic field of $50~\rm \mu T$ is cancelled out to better than a few $\mu$T using three mutually perpendicular sets of compensation coils. Thus in a ZF-$\mu$SR experiment, the polarization of incoming muons is measured in the sample in the absence of an external magnetic field. The $\mu$SR data were analyzed using the analysis software WiMDA.\cite{Pratt2000}

\small 
\begin{table*} [tbp] \caption{\label{Composition} Compositions of Li$_{1-x}$Fe$_x$(OH)Fe$_{1-y}$Se and lattice parameters for the samples in this study.}
	\begin{ruledtabular}
		\begin{tabular} {c c c c c c c}
			Label & $x$ & $y$ & $a (\rm \AA)$ & $c(\rm \AA)$ & Volume(\rm \AA$^3$) & Comments \\ 
			\hline NonSC1 & $0.203(8)$ & $0.156(7)$ & $3.8169(1)$ & $9.1744(6)$  & $133.665(1)$  & \\ 
			 NonSC2 & $0.207(2)$ & $0.129(1)$ & $3.8196(2)$ & $9.1706(1)$  & $133.793(2)$  &  \\
			 NonSC3 & $0.228(2)$ & $0.078(3)$ & $3.8054(1)$ & $9.2014(4)$  & $133.243(1)$  &  \\
			 SC1 & $0.162(2)$ & $0.011(4)$ & $3.7729(1)$ & $9.3590(2)$  & $133.228(2)$  &  \\
			 SC2 & $0.164(2)$ & $0.010(3)$ & $3.7808(1)$ & $9.3002(2)$  & $132.947(1)$ & Deuterated \\
		\end{tabular} 
	\end{ruledtabular}
\end{table*}
\normalsize

The precise compositions of each sample are detailed in Table \ref{Composition} as well as lattice parameters where $a$ and $c$ are the intralayer and interlayer parameters respectively.  SC2 was the subject of a recent neutron spectroscopy study.\cite{Davies2016}  Immediately noteworthy is the Fe occupation of the hydroxide layer which takes a  similar value for all five samples of $x \approx 0.2$ for non-superconductors and $x \approx 0.16$ for superconductors.  In contrast, the value of $y$ varies from approximately $0.01 \-- 0.16$.  This agrees with the $x$ and $y$ values of other published work on these compounds\cite{Pachmayr2015, Lu2014, Khasanov2016, Wu2015} and the required $x$ thought to be needed for structural stability.\cite{Chen2016}  Both SC1 and SC2 show slightly lower $x$ due to the post-synthetic lithiation they underwent which displaced hydroxide layer Fe to the FeSe layer, filling iron vacancies.

\section{Hydroxide Layer}

\begin{figure} [tbp]
	\includegraphics[width = .6\columnwidth]{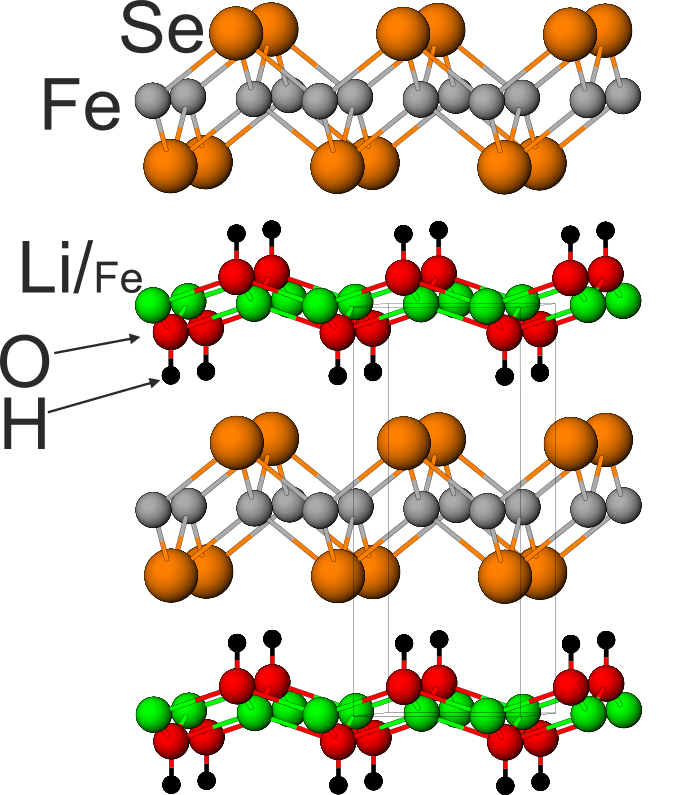}%
	\caption{Layer structure of Li$_{1-x}$Fe$_x$(OH)Fe$_{1-y}$Se.  The unit cell is also highlighted.\label{Struct}}
\end{figure}

\begin{figure} [tbp]
	\includegraphics[width = 0.8\columnwidth]{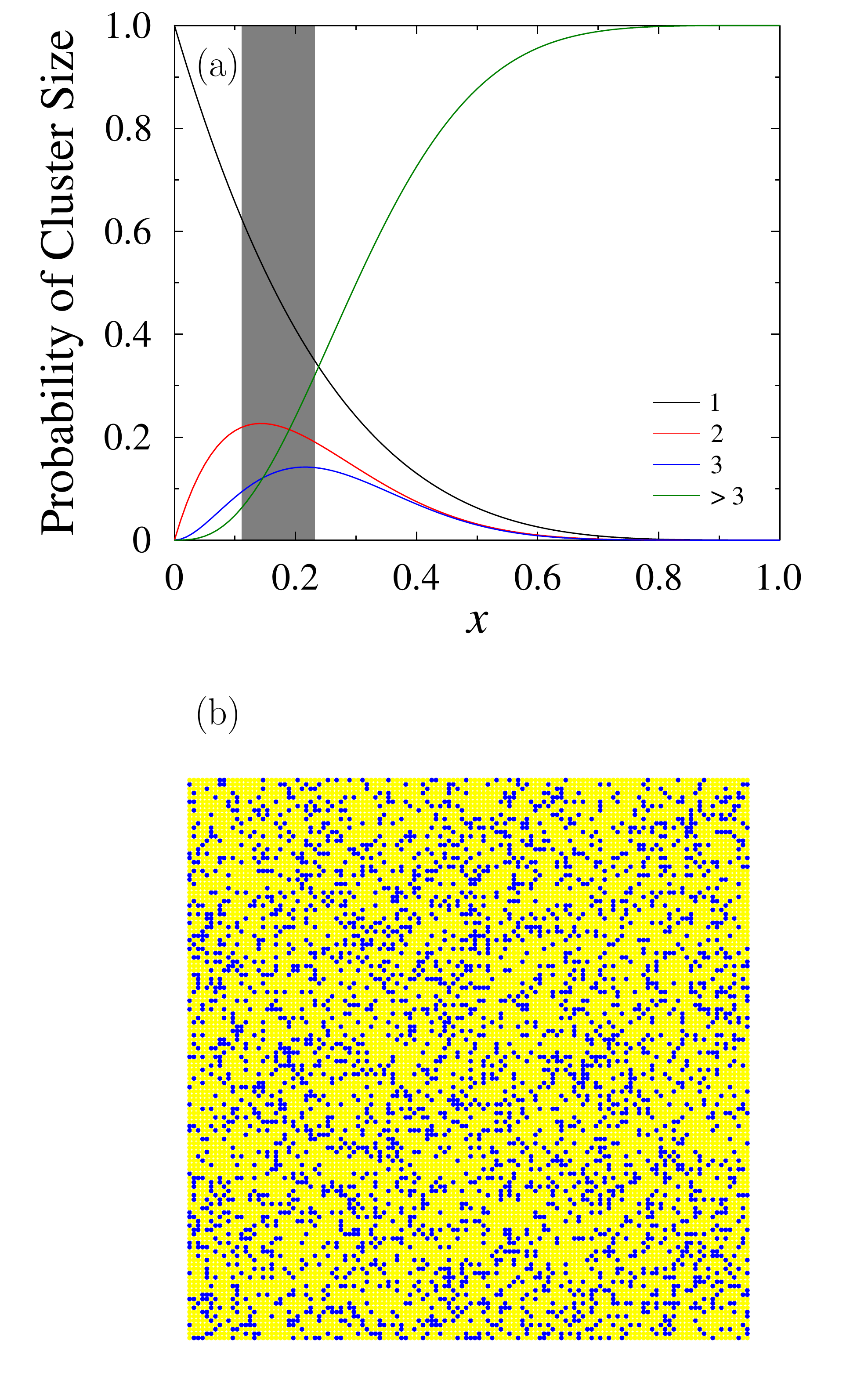}%
	\caption{(a) Fraction of Fe in the hydroxide layer in clusters of 1, 2, 3 and more than 3 Fe ions given that the initially considered ion is iron.  Distribution of Fe in the layer has been assumed random such that the probability of a hydroxide site containing Fe is equal to $x$.  Only nearest neighbours have been considered as constituting a cluster.  The region of $x$ spanned in this study is highlighted in grey.  (b) Graphical representation of the 2-D hydroxide layer with $x = 0.2$, Li = yellow and Fe = blue as shown in Ref.~\onlinecite{Woodruff2016}.\label{Percolation}}
\end{figure}

The hydroxide layer is comprised of a 2-D network of metal tetrahedra,
M(OH)$_4$, compressed along the stacking direction as shown in
Fig.~\ref{Struct}.  The resultant metal lattice forms two-dimensional
square sheets.  With an Fe occupancy of $x \approx 0.2$ in a Li
matrix, and no structural evidence for Li/Fe long range
order,\cite{Woodruff2016,Lynn2015} we can calculate the probability of
a specific Fe ion being isolated, part of a dimer, part of a trimer,
etc, under the assumption that the Li and Fe ions are randomly
distributed.  These probabilities are plotted as a function of $x$ in
Fig.~\ref{Percolation}(a).  This demonstrates that in Fe
concentrations ($x\approx0.16\--0.23$) under consideration the
majority of Fe ions exist as isolated ions, though there are also
dimers and clusters of greater than three ions.  It should be
highlighted that Fig.~\ref{Percolation}(a) shows only probability of
cluster size for a particular iron ion and not the fraction of clusters of a particular size.  Thus, despite dimers and clusters of greater than $3$ Fe ions having similar probabilities for this range of $x$, there will be a greater number of dimers.

Fig.~\ref{Percolation}(b) shows a schematic of a hydroxide layer randomly populated with $x = 0.2$.  A large number of clusters of varying sizes may be seen with no long range network across the sample (if only nearest neighbour interactions are considered), consistent with $x$ being well below the percolation threshold ($x_{\rm c}=0.5927$) for a two-dimensional square lattice.\cite{Ziff1992,StaufferBook}  Thus long range magnetic order sustained by nearest-neighbour interactions would not be expected.  Even taking into account longer range interactions, the competition that would arise between clusters due to differing cluster size, interaction type (ferromagnetic or antiferromagnetic), and possible anisotropies in each cluster would probably lead to a more frustrated ground state than simple long range magnetic order.  This analysis supports a glass-like magnetic interpretation of the hydroxide layer which has recently been proposed,\cite{Woodruff2016} whereas long-range order would be highly unlikely.

\section{D.c.\ Verification}

\begin{figure} [tbp]
	\includegraphics[width = 0.8\columnwidth]{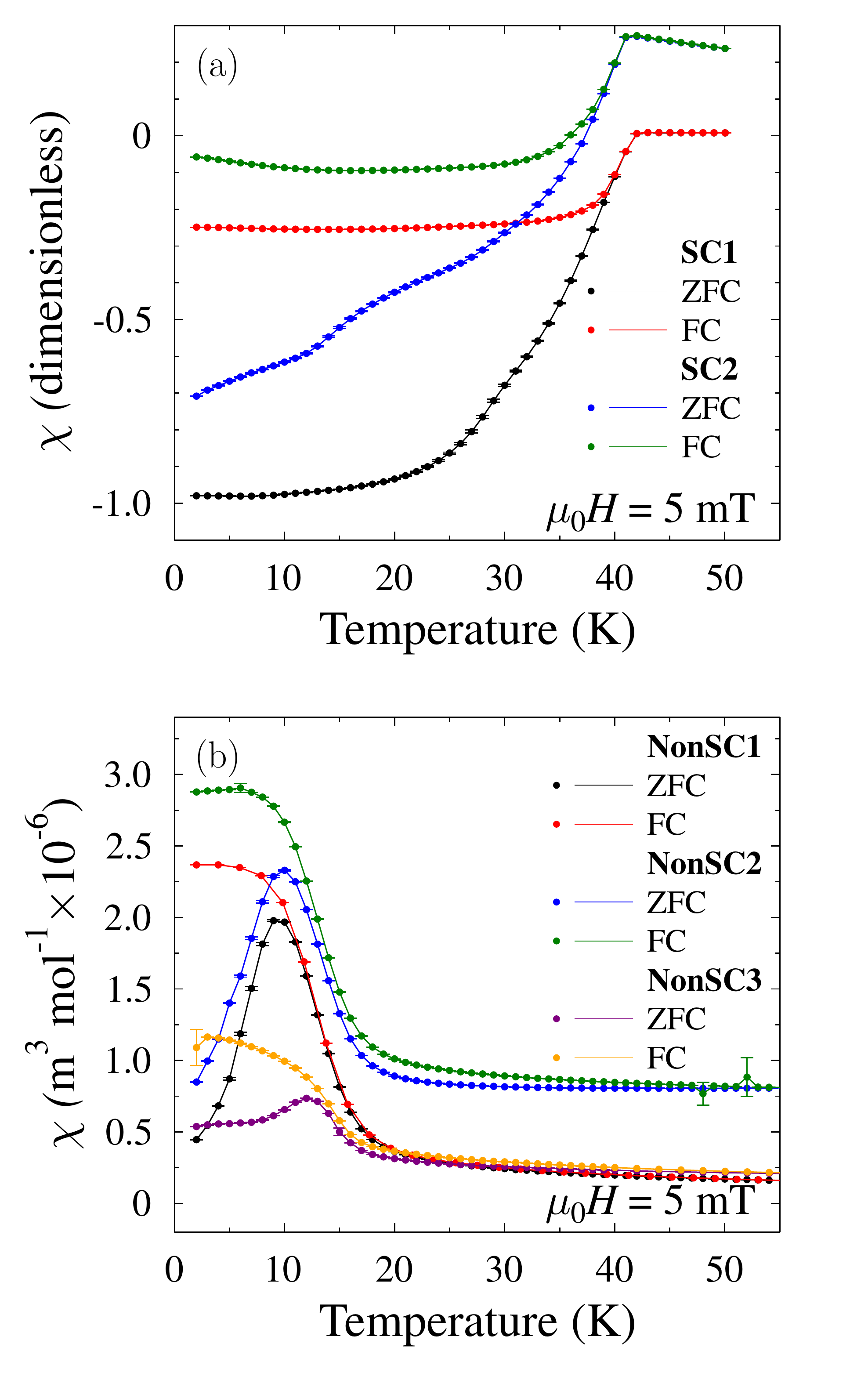}%
	\caption{(a) Magnetic susceptibility, $\chi$, of superconducting samples SC1 and SC2 measured in a d.c.\ magnetic field of $\mu_{\rm 0}H = 5~\rm mT$.  While an exact demagnetising factor is unknown an estimate of $\approx 0.4$ has been made by referring to the demagnetising factor of spherical particles in a cuboid container spanning values of $0.33\--0.43 $.\cite{Bjork2013}  (b) Magnetic susceptibility, $\chi$, of non-superconducting samples NonSC1, NonSC2 and NonSC3 measured in a d.c.\ magnetic field of $\mu_{\rm 0}H = 5~\rm mT$.  Due to smaller signal no demagnetising factor correction has been performed.\label{DCSusceptFig}}
\end{figure}

Results of initial d.c.\ magnetic characterisation are shown in Figs.~\ref{DCSusceptFig}(a) and (b).  SC1 and SC2 both show a strong superconducting transition with a $T_{\rm c}$ of approximately $40~\rm K$.  NonSC1, NonSC2 and NonSC3 show no superconducting signature but instead a ZFC peak (henceforth referred to as $T_{\rm p}$) in the range of $9$ to $12~\rm K$ accompanied by a divergence of ZFC and FC sweeps.  Both SC1 and SC2 show high diamagnetic shielding at low temperatures.  These plots show that both are high $T_{\rm c}$ FeSe superconductors with high volume fractions.

\begin{figure} [tbp]
	\includegraphics[width = 0.8\columnwidth]{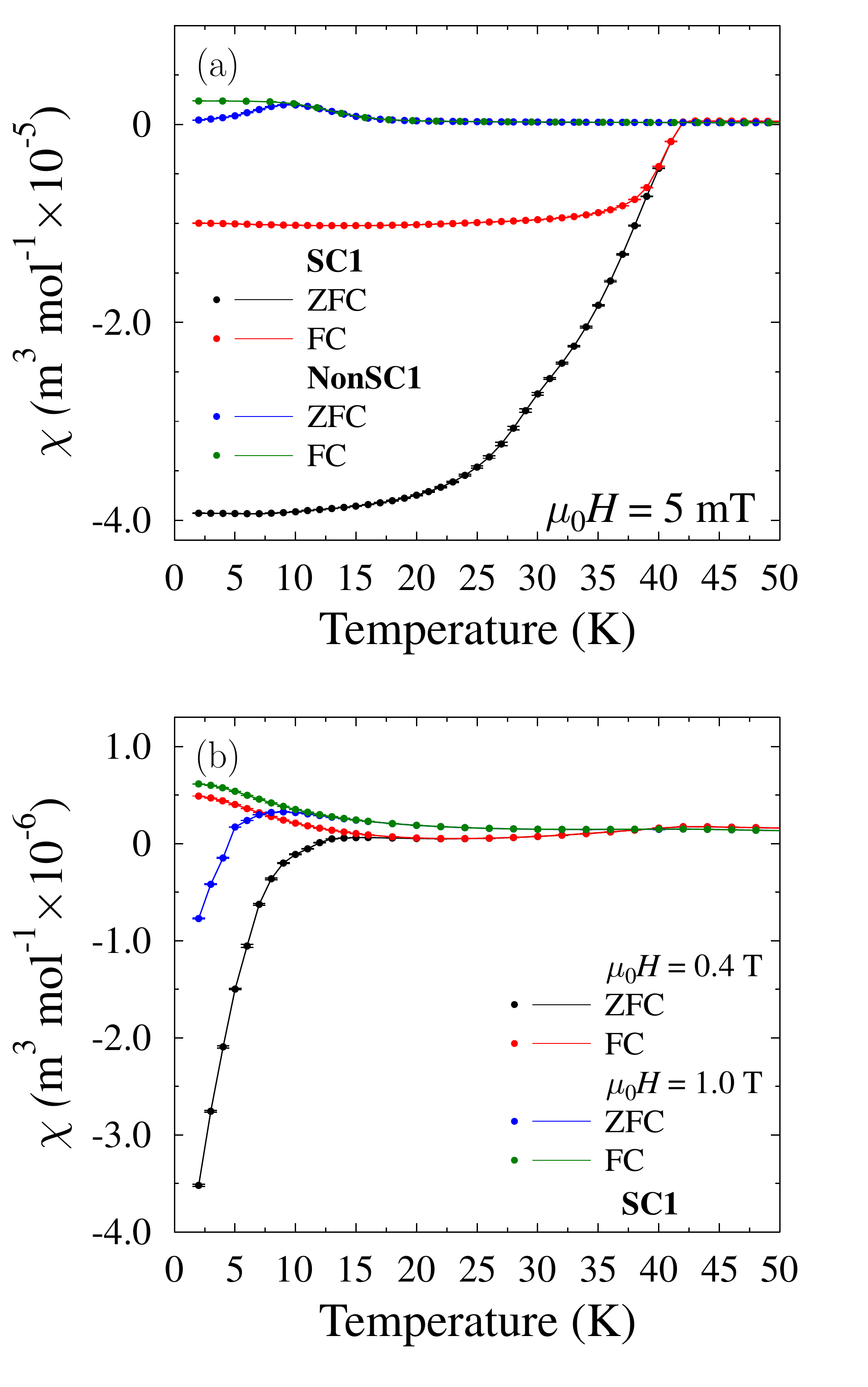}%
	\caption{(a) Magnetic susceptibility, $\chi$, of superconducting samples SC1 and NonSC1 measured in a d.c.\ magnetic field of $\mu_{\rm 0}H = 5~\rm mT$ from Fig. \ref{DCSusceptFig}(a) and (b) plotted with the same units for comparison.  (b) Magnetic susceptibility, $\chi$, of superconducting sample SC1 measured in d.c.\ magnetic fields of $\mu_{\rm 0}H = 0.4~\rm T$ and $\mu_{\rm 0}H = 1.0~\rm T$.\label{DCSusceptFig_Extra}}
\end{figure}

The ZFC peaks at $\approx 10~\rm K$ in the non-superconducting samples as well as the divergence between ZFC and FC sweeps suggests the development of some type of at least local magnetic order but is not, in itself, sufficient to distinguish between ferromagnetic or glass-like behaviour.\cite{Binder1986}  A ZFC peak in this temperature region has been detected in superconducting variants of these compounds which was interpreted as long range ferromagnetic order.\cite{Pachmayr2015}  However, a ZFC peak is not present in SC1 or SC2 nor in other members of this family of compounds.\cite{Lu2014,Sun2015} This suggests that the peak is fairly sensitive to $x$ as it has been detected in a previously reported  member of this family which superconducts with $x \approx 0.2$.\cite{Pachmayr2015}  Indeed, due to the diamagnetic signal of superconductivity being of a much greater magnitude than that of the $T_{\rm p}$ feature a peak may be too relatively weak to be detected for SC1 and SC2.  This is demonstrated in Fig.~\ref{DCSusceptFig_Extra}(a) in which the d.c. susceptibility sweeps of SC1 and NonSC1 have been plotted in the same units.  The low temperature feature of NonSC1 is very small relative to the superconducting signal of SC1.  If the magnitude of this was reduced due to the lower $x$ of the superconductors and it was superimposed on top of the superconducting SC1 it would likely be difficult to detect.  Furthermore, a superconducting member of this family lacking this feature in a magnetic field of $\mu_0H=1~\rm mT$ was shown to display very similar susceptibility to the non-superconductors with a ZFC peak at $8.6~\rm K$ accompanied by a divergence of ZFC and FC sweeps when measured in a high magnetic field of $\mu_0H=1.0~\rm T$.\cite{Lu2014}  This was explained by the $1.0~\rm T$ field suppressing superconductivity and allowing only the hydroxide layer magnetism to be observed, which was interpreted in Ref.~\onlinecite{Lu2014} as canted antiferromagnetism.

A test to reproduce this effect for SC1 is shown in Fig.~\ref{DCSusceptFig_Extra}(b).  While not completely suppressing the superconductivity, partial suppression has occurred with a very asymmetric peak at approximately $T_{\rm p}$ which moves to a lower temperature with a higher magnetic field accompanied by a divergence between ZFC and FC sweeps.  These features fit a spin glass description and Fig.~\ref{DCSusceptFig_Extra}(b) may be explained thus: for the ZFC sweep the moments of the hydroxide layer at low temperatures are frozen in random orientations resulting in no net magnetisation and only partial diamagnetic shielding from the superconductivity being present.  On warming the system and passing through $T_{\rm p}$ (which we identify as the spin glass freezing temperature, $T_{\rm f}$) the moments become unfrozen and align with the applied field.  This results in an enhancement of the magnetic field experienced by the sample and further suppresses the superconductivity.  It is not completely suppressed so an anomaly at $T_{\rm c}$ can still be detected, resulting in a small minimum occurring between $T_{\rm p}$ and $T_{\rm c}$.  For the FC sweep, due to cooling in a magnetic field the moments in the hydroxide layer become frozen and aligned with the magnetic field resulting in the near complete suppression of superconductivity even at $2~\rm K$ (though a small anomaly persists at $T_{\rm c}$).

\section{A.c.\ Susceptibility}

\begin{figure*} [tbp]
	\centering
	\includegraphics[width = 1.6\columnwidth]{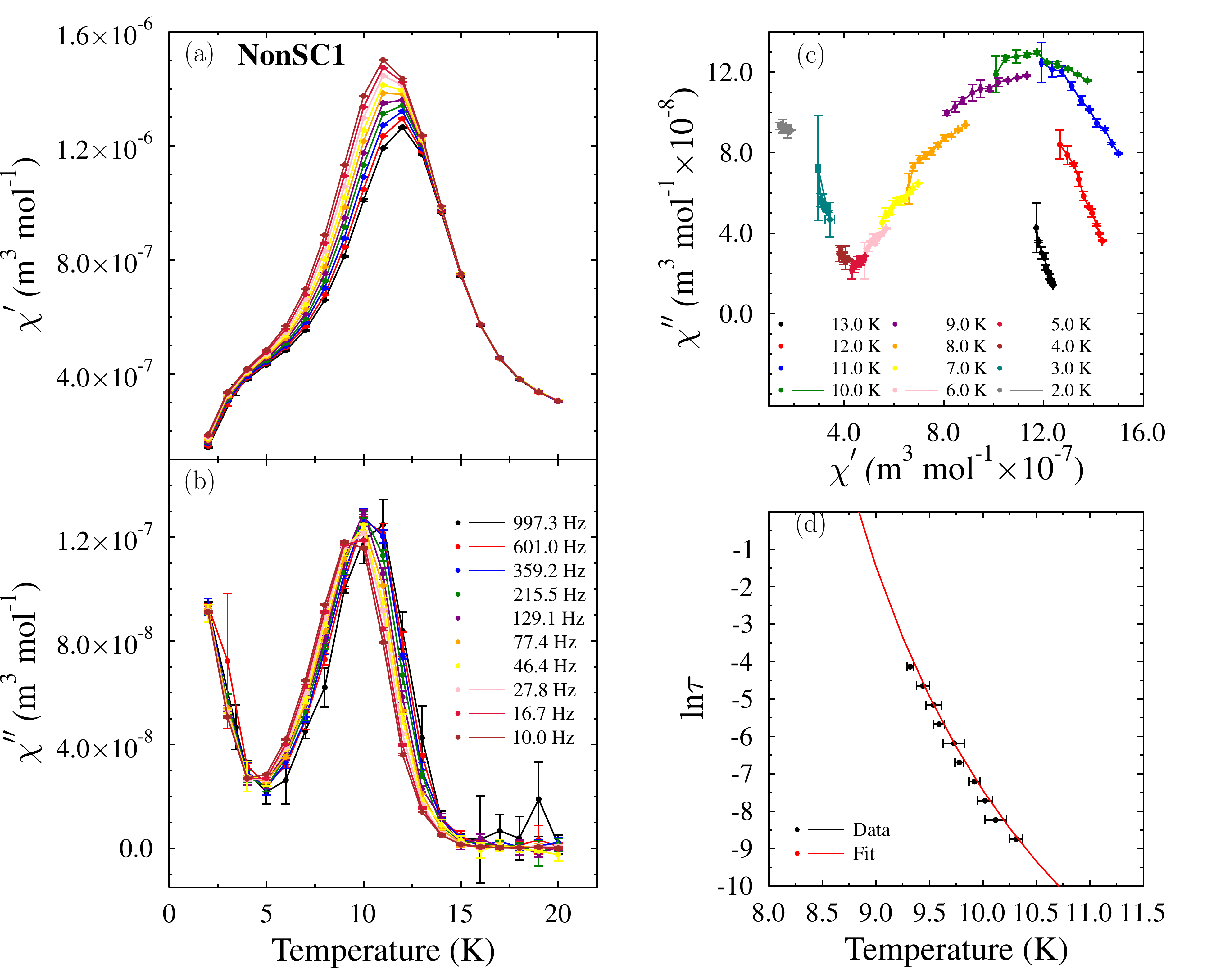}%
	\caption{(a) and (b) show a.c.\ susceptibility of NonSC1 as measured at $\mu_{0}H_{\rm d.c.}= 0~\rm T$ with an a.c.\ magnetic field amplitude of $\mu_{0}H_{\rm a.c.}= 0.4~\rm mT$.  (a) shows the real/in-phase parts of susceptibility, $\chi^{\prime}$ while (b) shows the imaginary/out-of-phase parts of susceptibility, $\chi^{\prime\prime}$.  This was previously published in Ref. \onlinecite{Woodruff2016}.  (c) Cole-Cole plot of ac susceptibility shown in (a) and (b).  (d) Temperature of $\chi^{\prime\prime}$ peaks from (b) plotted with fit to Volger-Fulcher law where $\tau = 1/\omega$. These were determined by Gaussian fits to the peaks from $7 \-- 13~\rm K$ .\label{acSusceptNonSC}}
\end{figure*}

The measurement of a.c.\ susceptibility was employed in a previous study\cite{Woodruff2016} to elucidate the nature of the magnetic properties of NonSC1 with the results from this study shown in Figs.~\ref{acSusceptNonSC}(a) and (b) as the real/in-phase part of the susceptibility ($\chi^{\prime}$) and the imaginary/out-of-phase part of the susceptibility ($\chi^{\prime\prime}$) respectively.  The frequency dependence of $\chi^{\prime}$ develops only below $T_{\rm p}$ and is accompanied by a set of frequency dependent peaks in $\chi^{\prime\prime}$.  This bears similarities to that of slow relaxing single-molecule magnets\cite{GatteschiBook,Balanda2013} and spin glasses.\cite{Balanda2013,Binder1986,MydoshBook}  The characteristic value $\Delta T_{\rm f}/[T_{\rm f}\Delta (\ln(\omega))]$ (where $T_{\rm f}$ is the freezing temperature corresponding to peaks in $\chi^{\prime}$ and $\omega$ is the frequency of the applied a.c.\ magnetic field) was found to be $0.016(1)$,\cite{Woodruff2016} consistent with a spin glass.\cite{GatteschiBook}

To understand this behaviour more deeply, we show the Cole-Cole plot for $\chi^{\prime\prime}$ v. $\chi^\prime$ in Fig.~\ref{acSusceptNonSC}(c) which for a system dominated by a single relaxation time (such as a single molecule magnet) would yield isothermal arcs.\cite{GatteschiBook}  Instead for NonSC1 we find what appear to be very small slices of arcs in the region of $T_{\rm p}$.  This provides evidence for a system containing slow magnetic relaxation with a spread of relaxation times, functioning to distort the arcs from the ideal semicircle of a single relaxation time,\cite{GatteschiBook, Binder1986} similar to that observed in spin glasses.\cite{MydoshBook,Binder1986}

Further anomalies not observed in d.c.\ susceptibility measurements can be identified in Fig.~\ref{acSusceptNonSC}(a) and (b) as a closing of frequency dependence with a shoulder in $\chi^{\prime}$ at $\approx 4~\rm K$ and upturn in $\chi^{\prime\prime}$ at $\approx 5~\rm K$.  This manifests in the Cole-Cole plot as a set of isothermal points that span a very narrow region of $\chi^\prime$.  Though the frequency range available limits investigation, the $\chi^{\prime\prime}$ upturn may suggest the presence of a second, unknown relaxation process becoming important at lower temperatures.

\begin{figure*} [tbp]
	\includegraphics[width = 1.6\columnwidth]{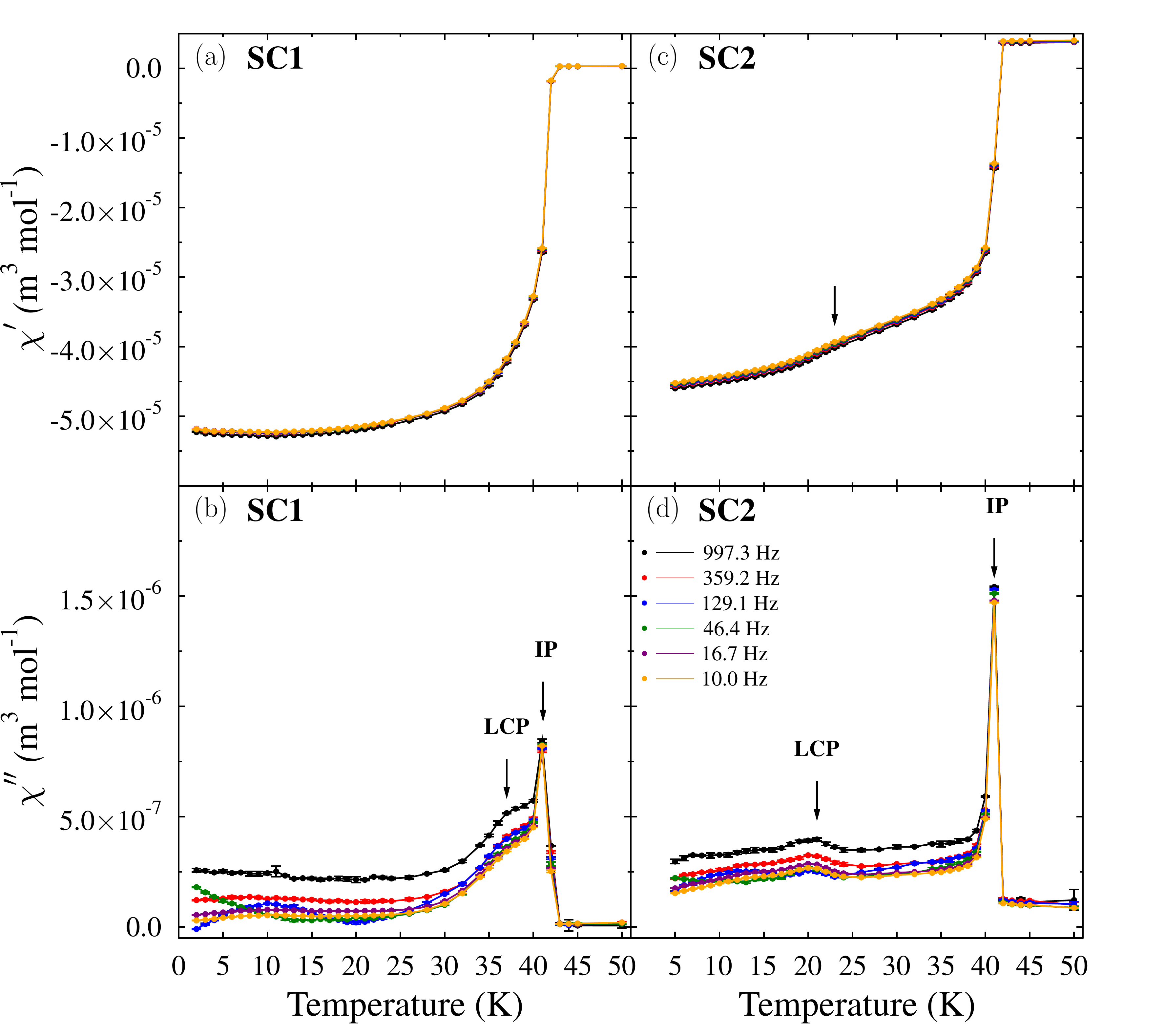}%
	\caption{A.c. susceptibility of superconducting compounds at $\mu_{0}H_{\rm d.c.}= 0~\rm T$ with an a.c. magnetic field amplitude of $\mu_{0}H_{\rm a.c.}= 0.4~\rm mT$.  (a) and (b) show the real/in-phase and imaginary/out-of-phase parts or susceptibility, $\chi^{\prime}$ and $\chi^{\prime\prime}$, of SC1.  (c) and (d) show the real/in-phase and imaginary/out-of-phase susceptibility, $\chi^{\prime}$ and $\chi^{\prime\prime}$, of SC2.\label{acSusceptACSC}}
\end{figure*}

Analysis of the peaks of $\chi^{\prime\prime}$ was performed using the Arrhenius law,

\begin{equation}
	\tau = \tau_0 \exp\left[\frac{E_{\rm a}}{k_{\rm B}T}\right],
\end{equation}
and the Vogel-Fulcher law,

\begin{equation}
\tau = \tau_0 \exp\left[\frac{E_{\rm a}}{k_{\rm B}(T-T_0)}\right],
\end{equation}
where $\tau = 1/\omega$, $\tau_0$ is the characteristic relaxation time, $E_{\rm a}$ is the activation energy for the relaxation process, $T$ is the temperature of the peaks in $\chi^{\prime\prime}$ and $T_0$ is the characteristic temperature.\cite{Tholence1980,GatteschiBook,Balanda2013,Binder1986,Shtrikman1981,MydoshBook}  The Arrenhius fit produced unphysical values (e.g.\ $E_{\rm a} = 470\pm 10~\rm K$) but the Vogel-Fulcher fit shown in Fig.~\ref{acSusceptNonSC}(d) yields $T_0 = 6.4\pm 1.2~\rm K$, $\ln \tau_0 = -23 \pm 6$ and $E_{\rm a} = 60 \pm 40~\rm K$, the large uncertainty in parameter values resulting from the narrow temperature range over which $\chi^{\prime\prime}$ peaks exist.  Unphysical parameter values from an Arrhenius fit are often found in a glass-like system\cite{Tholence1980,Balanda2013,Binder1986,Shtrikman1981} in which the non-interacting moments described by the Arrhenius law (as in single molecule magnets\cite{GatteschiBook}) are not present.  The Vogel-Fulcher law is typically used as an improvement over the Arrhenius law for spin glasses\cite{Tholence1980,Balanda2013,Binder1986,Shtrikman1981}.    This demonstrates that a spin glass description of the slow magnetic relaxation of the system is reasonable with slight deviations from the Vogel-Fulcher law unsurprising as it is only a phenomenological description.\cite{Binder1986}

A.c.\ susceptibility measurements of the superconductors SC1 and SC2 are shown in Fig.~\ref{acSusceptACSC}.  $\chi^{\prime}$ of each compound (Fig.~\ref{acSusceptACSC}(a) and (c)) both show strong superconducting transitions though a small anomaly can be detected as a broad hump for SC2 in the region of $25~\rm K$ (arrow in Fig.~\ref{acSusceptACSC}(c)).  The out-of-phase susceptibility (Fig.~\ref{acSusceptACSC}(b) and (d)) of SC1 and SC2 appear very similar with a small non-zero $\chi^{\prime\prime}$ above $T_{\rm c}$, a sharp seemingly frequency independent peak at $T_{\rm c}$ and an approximately steady, non-zero $\chi^{\prime\prime}$ at all lower temperatures.  SC1 shows a shoulder just below $T_{\rm c}$ while SC2 shows a bump at $\approx 20~\rm K$ (both indicated in Fig.~\ref{acSusceptACSC}(b) and (d)).

Though one might expect the dissipative $\chi^{\prime\prime}$ signal to be zero in a superconductor, dissipation can occur due to vortices.\cite{Nikolo1994}  In the studied systems should the hydroxide layer magnetism and superconductivity be well separated (as we have assumed since they originate from different layers) the frequency dependent peaks of $\chi^{\prime\prime}$ observed in NonSC1 would be superimposed upon a superconducting $\chi^{\prime\prime}$ only non-zero in the region of $T_{\rm c}$.  Due to the significant temperature separation of $T_{\rm p}$ and $T_{\rm c}$ this would result in two distinct peaks at each frequency at the low magnetic fields considered.

This is not observed in Fig.~\ref{acSusceptACSC}(b) or (d), perhaps due to the large non-zero $\chi^{\prime\prime}$ of each superconductor masking low temperature peaks or the lower iron concentration of the hydroxide layer of the superconductors.  Two sets of peaks in $\chi^{\prime\prime}$ are present for SC1 and SC2 which are identified as the intrinsic peaks (IP) corresponding to the superconducting transition of the samples and lower coupling peaks (LCP) corresponding to the superconducting transition of the powder grain boundaries.\cite{Nikolo1989,Nikolo1994}  A bulge in $\chi^{\prime}$ at a slightly higher temperature accompanies the LCP and this is highlighted by an arrow for SC2.  For SC1 this is hidden by the rapid drop of $\chi^\prime$ at $T_{\rm c}$.

\section{Heat Capacity}

\begin{figure*}[htbp]
	\includegraphics[width = 1.6\columnwidth]{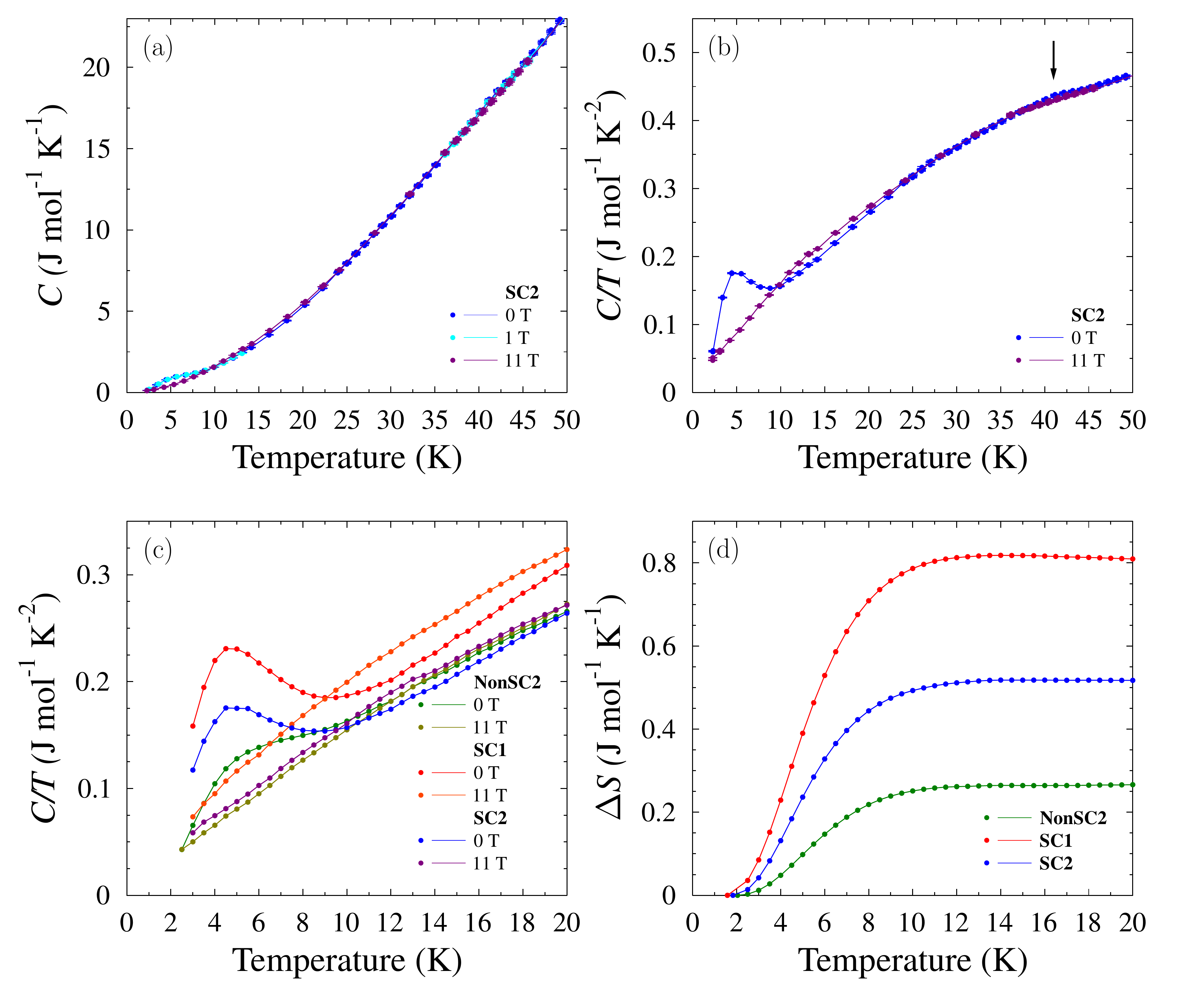}%
	\caption{(a) Heat capacity of SC2 measured at various static magnetic fields ($\mu_{0}H_{\rm d.c.}= 0, 1$ and $11~\rm T$).  No anomaly can be observed at $T_{\rm c}$.  (b) $C/T$ v. $T$ of SC2 at magnetic fields of $\mu_{0}H_{\rm d.c.}= 0$ and $11~\rm T$.  An anomaly at $T_{\rm c}$ may exist in the form of a very subtle bulge indicated by an arrow.  (c) $C/T$ v. $T$ of NonSC2, SC1 and SC2 at magnetic fields of $\mu_{0}H_{\rm dc}= 0$ and $11~\rm T$.  (d) Entropy change for NonSC2, SC1 and SC2 obtained as detailed in text.\label{HCFig}}
\end{figure*}
Heat capacity measurements were performed on each of NonSC1, NonSC2,
SC1 and SC2.  Both superconductors showed similar behaviour and,
therefore, the full region of interest is reported for SC2 only in
Fig.~\ref{HCFig}(a) and (b).  Unexpectedly, no feature can be
distinguished in heat capacity at $T_{\rm c}$ with no visible change
with the application of $11~\rm T$.  Instead, the only feature present
is a low temperature bump with a peak at $\approx 5~\rm K$.  This bump
is suppressed in magnitude, broadened and pushed to higher
temperatures in high magnetic fields.  When examining $C/T$ the low
temperature feature is accentuated whereas a $T_{\rm c}$ anomaly may
perhaps be identified on top of the lattice contribution (highlighted
by an arrow in Fig. \ref{HCFig}(b)).  However, this is weak enough
that it may be a measurement artefact.  The lack of a distinct feature
at $T_{\rm c}$ for these compounds is not entirely unexpected.
Elemental FeSe shows only a small heat capacity anomaly at $T_{\rm
  c}$\cite{Bohmer2015} and investigations of
Ba$_{0.8}$Fe$_2$Se$_2$\cite{Ying2012} and LiFeAs\cite{Baker2009} show
the difficulty of detecting anomalies at $T_{\rm c}$ in these
compounds.  Furthermore, the Sommerfeld coefficient (obtained from
third order polynomial fits to $C/T$ v. $T$ from $13\--50~\rm K$ for
$0~\rm T$ data) for SC1 and SC2 was found to be $\gamma = 3(6)$ and
$0.0(7)~\rm mJ~mol^{-1}~K^{-2}$ for each compound respectively.  Due
to the low temperature anomaly limiting the lower bounds of the fits
and the temperature region lying within the superconducting regime,
exact determination of $\gamma$ is difficult with uncertainties likely
larger than those quoted from polynomial fits.  An upper bound may be
placed on $\gamma$ of $\approx 50~\rm mJ~mol^{-1}~K^{-2}$ by
projecting $0~\rm T$ data of Fig.~\ref{HCFig}(c) to $T=0$~K.  These suggest SC1
and SC2 show a very small electronic contribution to heat capacity
similar to LiFeAs (with $\gamma \approx 23.3(5)~\rm
mJ~mol^{-1}~K^{-2}$) which also displayed a weak discontinuity of heat
capacity at $T_{\rm c}$.\cite{Baker2009} As such, the lattice
contributions may be much greater which, combined with the smearing
out of features due to the samples being pressed pellets, hide an
anomaly at $T_{\rm c}$.  Our results are in agreement with those
reported in an independent investigation\cite{Lu2014} in which heat
capacity was measured in the range $2\-- 20~\rm K$ for a
superconducting sample with a similar composition.

Fig.~\ref{HCFig}(c) shows $C/T$ in the low temperature region for NonSC2, SC1 and SC2 with each showing the same anomaly at $5\--6~\rm K$ that is suppressed in magnitude at higher applied magnetic fields.  The data shown are interpolated results of several sweeps which facilitates the subtractions needed for the calculation of entropy.  Slight variations between samples may be attributed to small errors when determining pellet masses or background contributions with the larger high temperature values of SC1 perhaps indicating a more homogeneous sample.  SC1 did show the largest superconducting volume fraction in Fig.~\ref{DCSusceptFig}(a).  As these peaks extend up to temperatures of $\approx T_{\rm p}$ they can be identified as originating from the same feature.  Both SC1 and SC2 (with $x \approx 0.16$) show similarly sharp peaks while the peak in NonSC2 ($x\approx0.21$) is much less prominent appearing more as a shoulder.  However, a superconducting member of this family with $x=0.2$ also showed a sharp peak in $C/T$ at this temperature suggesting that this small variance of $x$ may not be the controlling factor.  Therefore, it appears the factors that favour superconductivity in these compounds also favour an enhanced entropy change due to the magnetic state of the hydroxide layers.

The suppression of the low temperature peaks in a magnetic field of $11~\rm T$ is reminiscent of spin glasses\cite{Binder1986} though it occurs over a narrower temperature range than expected.\cite{MydoshBook,Wenger1976}   To calculate the magnetic contribution $\Delta C/T$ a third order polynomial was fit to the $0~\rm T$ $C/T$ data for each sample from $13\--50~\rm K$ and subtracted from each $0~\rm T$ sweep from $2\--20~\rm K$.  $\Delta C/T$ for each sample was linearly extrapolated to zero using the two lowest temperature points for each sample.  This was subsequently integrated to yield the change in entropy through the transition, $\Delta S$, as shown in Fig.~\ref{HCFig}(d).  While these values may be considered as lower bounds on $\Delta S$ since they are limited by the minimum temperature accessible via experiment, they are considerably lower than expected.  Defining the change in entropy through the magnetic transition as $\Delta S = xR\ln [2J+1]$ (where $R$ is the gas constant, $x$ the Fe concentration of the hydroxide layer and $J$ is the effective spin of the magnetic moments in the system\cite{Wenger1976}) yields the $J$ values in Table.~\ref{EntopyResults}.  While uncertainties may be slightly underestimated due to the fitting procedure employed to find $\Delta C/T$, these values of $J$ all lie below the value for one free electron ($1/2$).  Furthermore, the spin expected from crystal field theory for the Fe$^{2+}$ ions in a tetrahedral coordination is $2$ and DFT calculations predict a moment of $\approx 3.5~\mu_{\rm B}$ for hydroxide layer Fe.\cite{Chen2016,Liu2017}  This analysis therefore lends support to the suggestion that the hydroxide layer has glassy magnetic character as such spin underestimation has been previously observed in spin glasses.\cite{Wenger1976}

Thus, in contrast with a previous interpretation of these low temperature heat capacity peaks as evidence of canted antiferromagnetism,\cite{Lu2014} they suggest more probably a magnetic state common to both superconducting and non-superconducting compounds resembling a spin glass.

\small 
\begin{table} [tbp] \caption{\label{EntopyResults} Entropy change from low temperature transition and predicted moments of the hydroxide layer from Fig. \ref{HCFig}(d).  $\Delta S$ has been obtained from linear fits between the temperatures of $13\--20~\rm K$ with gradient fixed to $0$.}
	\begin{ruledtabular}
		\begin{tabular}{c | c c }
			 & $\Delta S$ (J mol$^{-1}$ K$^{-1}$) & Spin ($J$) \\ 
			\hline NonSC2 & $0.2646(2)$& $0.083(3)$  \\ 
			SC1 & $0.7856(9)$ & $0.396(8)$  \\ 
			SC2 & $0.5166(3)$ & $0.230(7)$  \\ 
			
		\end{tabular} 
	\end{ruledtabular}
\end{table}
\normalsize

\section{ZF-$\mu$SR}

\begin{figure} [htbp]
	\includegraphics[width = 0.8\columnwidth]{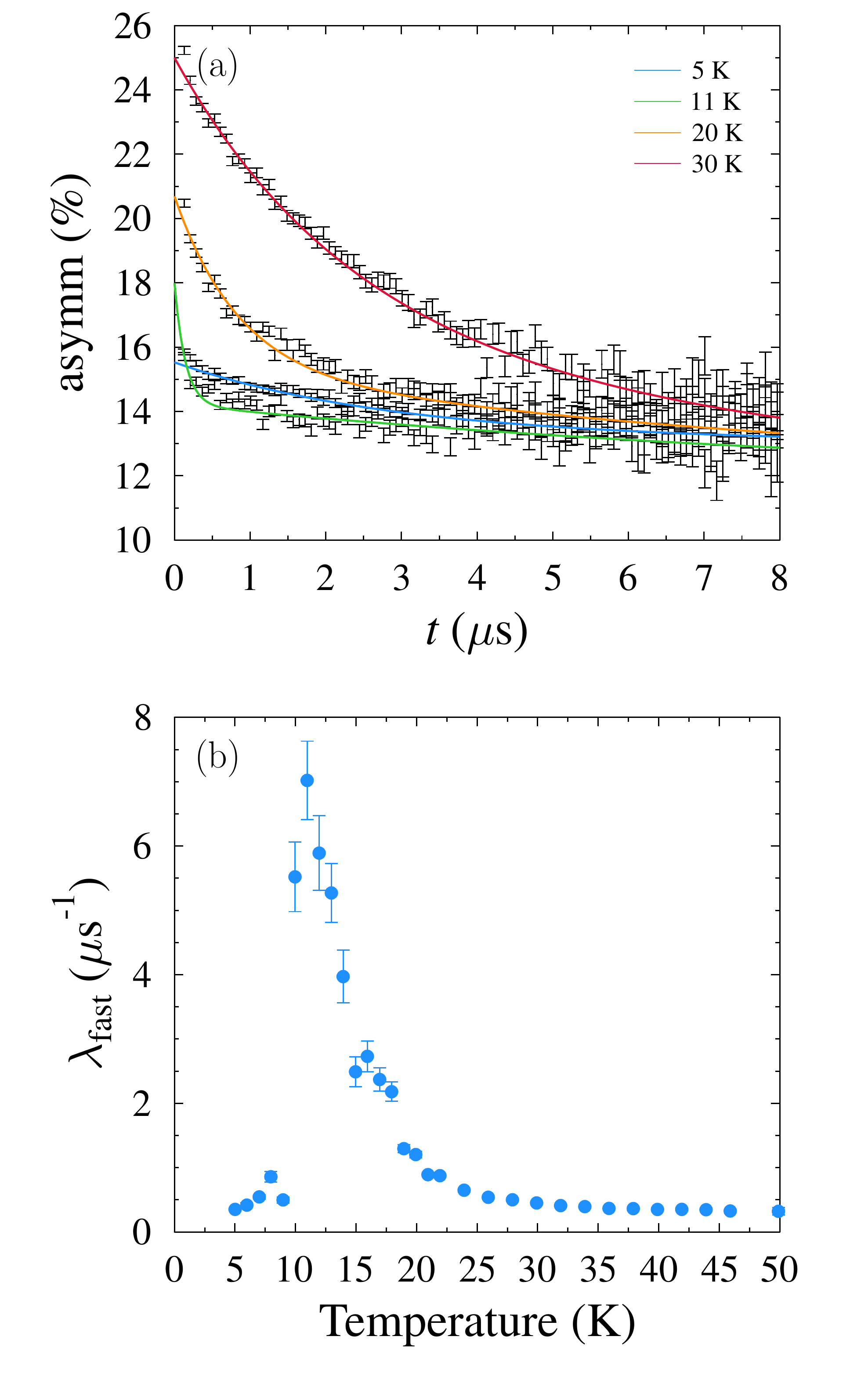}%
	\caption{(a) Asymmetry of muons at various temperatures as measured in NonSC3.  (b) Fast relaxation rates, $\lambda_{\rm fast}$, of muons at various temperatures from fits to data as shown in (a).\label{muSRRes}}
\end{figure}

Example ZF-$\mu$SR spectra for NonSC3 are shown in Fig~\ref{muSRRes}(a). No Kubo-Toyabe-like relaxation or oscillatory signal was observed at any measured temperature in the forward-backward asymmetry spectra.  Thus no evidence for long range order or static magnetism was found, though the pulsed structure of the ISIS beam precludes an observation of precession signals faster than about 10~MHz.

The data were fitted with a two-component exponential relaxation function
\begin{equation} \label{Lor}
A(t)=A_{1}\exp\left(-\lambda_{\rm fast} t\right) + A_2\exp\left(-\lambda_{\rm slow} t\right),
\end{equation}
where the total relaxing amplitude consists of a slowly-relaxing background with amplitude $A_{2}$ and relaxation rate $\lambda_{\rm slow}$, and a further signal with amplitude $A_1$ and a faster relaxation rate $\lambda_{\rm fast}$. Exponential relaxation corresponds to either dynamic moments with a single correlation time within the resolution of the spectrometer and an unknown field distribution,\cite{Khasanov2008} or a dilute distribution of static moments.\cite{Walstedt1974} The slow relaxation was found to be approximately temperature-independent (with $\lambda_{\rm slow} \approx 0.1~\mu\rm{s}^{-1}$). The faster relaxation shows strong temperature dependence between 5~K and 50~K (as shown in Fig.~\ref{muSRRes}(b)), increasing sharply at temperatures below 25~K before peaking at 11~K. This corresponds with the peaks observed in a.c.\ susceptibility and heat capacity and provides further evidence for this transition representing a glassy transition in which clusters of iron spins within the hydroxide layer start to become static. To understand this more fully we carried out simulations of the muon polarization resulting from low concentrations of static and dilute magnetic moments.  It was found that the observed relaxation could not be satisfactorily modelled by either static spins or dynamic moments with a single relaxation time, supporting the notion obtained from a.c.\ susceptibility that a range of relaxation times are required. This is consistent with glassy behavior, probably originating from a distribution of cluster sizes within the hydroxide layer.

\section{Conclusion}

In conclusion, the synthesis of superconducting and non-superconducting members of the  Li$_{1-x}$Fe$_x$(OH)Fe$_{1-y}$Se family of compounds has allowed a comparative study in order to identify the magnetic character of the hydroxide layer.  While d.c. magnetometry did reveal the presence of a magnetic phase at $\approx 10~\rm K$ in non-superconducting compounds which was intermittently observed in superconducting variants, both a.c.\ magnetometry and $\mu$SR measurements found this feature not to correspond to long range order.  Instead, this appeared consistent with slow relaxation of magnetisation such as that found in spin glasses containing frustration as can be visualised from a random distribution of Fe$^{2+}$ ions in the hydroxide layer.  These were found to correlate with a low temperature feature in heat capacity present in both superconducting and non-superconducting members of this family suggesting the hydroxide layer magnetism is present regardless of the presence of superconductivity.  Furthermore, the suppression of this feature with magnetic field and an estimate of the entropy change suggest an spin less than that calculated for an Fe$^{2+}$ tetrahedral complex of $J=2$.  This offers strong support for the glass-like nature of this magnetism i.e.\ interacting, disordered moments.

\section{Acknowledgements}

This work was supported by the U. K. Engineering and Physical Sciences
Reasearch Council (EPSRC) (Grant No. ER/M020517/1), the Leverhulme
Trust (Grant No. RPG-2014-221) and the Oxford Centre for Applied
Superconductivity.  Work at the Diamond Light Source was funded by
allocation EE13284.  We thank E. Suard for assistance on the D2B
beamline at the Institut Laue-Langevin.  C.\ V.\ T.\ would like to
thank D. Prabhakaran for experimental assistance and expertise and
both D.~Prabhakaran and N.~Davies for fruitful discussions.
F.\ K.\ K.\ K.\ thanks Lincoln College Oxford for a doctoral
studentship. Part of this work was performed at the UK Science and
Technology Facilities Council (STFC) ISIS facility, Rutherford
Appleton Laboratory.

\bibliographystyle{apsrev4-1}
\bibliography{LiOHFeSe_HydroxideMagnetism_Ref}
\end{document}